\documentclass[aps,showkeys,showpacs,superscriptaddress,twocolumn,floatfix]{revtex4}
\usepackage{graphicx}
\usepackage{psfrag}
\usepackage{amsmath}
\usepackage{amsfonts}
\usepackage{color}

\begin{document}

\title{Critical behaviour of the Ising $S=1/2$ and $S=1$ model on $(3,4,6,4)$ and $(3,3,3,3,6)$ Archimedean lattices}

\author{F. W. S. Lima}
\email{fwslima@gmail.com}
\affiliation{
Departamento de F\'{\i}sica,
Universidade Federal do Piau\'{\i},\\
57072-970 Teresina, Piau\'{\i}, Brazil
}

\author{J. Mostowicz}
\affiliation{
Faculty of Physics and Applied Computer Science,
AGH University of Science and Technology,\\
al. Mickiewicza 30, PL-30059 Krak\'ow, Poland
}

\author{K. Malarz}
\homepage{http://home.agh.edu.pl/malarz/}
\email{malarz@agh.edu.pl}
\affiliation{
Faculty of Physics and Applied Computer Science,
AGH University of Science and Technology,\\
al. Mickiewicza 30, PL-30059 Krak\'ow, Poland
}

\date{\today}

\begin{abstract} We investigate the critical properties of the Ising $S=1/2$ and $S=1$ model on $(3,4,6,4)$ and $(3^4,6)$ Archimedean lattices.
The system is studied through the extensive Monte Carlo simulations.
We calculate the critical temperature as well as the critical point exponents $\gamma/\nu$, $\beta/\nu$ and $\nu$ basing on finite size scaling analysis.
The calculated values of the critical temperature for $S=1$ are $k_BT_C/J=1.590(3)$ and $k_BT_C/J=2.100(4)$ for $(3,4,6,4)$ and $(3^4,6)$ Archimedean lattices, respectively.
The critical exponents $\beta/\nu$, $\gamma/\nu$ and $1/\nu$ for $S=1$ are $\beta/\nu=0.180(20)$, $\gamma/\nu=1.46(8)$ and $1/\nu=0.83(5)$ for $(3,4,6,4)$ and $0.103(8)$, $1.44(8)$ and $0.94(5)$ for $(3^4,6)$ Archimedean lattices.
Obtained results differ from the Ising $S=1/2$ model on $(3,4,6,4)$, $(3^4,6)$ and square lattice.
The evaluated effective dimensionality of the system for $S=1$ are $D_{\text{eff}}=1.82(4)$ for $(3,4,6,4)$ and $D_{\text{eff}}=1.64(5)$ for $(3^4,6)$.
\end{abstract}



\maketitle

\section{Introduction}

The Ising model \cite{ising1,ising2} remains probably the most cited model in statistical physics.
Today, the {\em ISI Web of Knowledge} abstracting and indexing service returns over eleven thousands records for the query on ``Ising'' for time span from 1996 to 2010.
For {\em Inspec} database (for years 1969-2010) this number is almost doubled and reaches 19 thousands for {\em Scopus} database (for data range 1960-2010).
The latter means that during the last half of century $\approx 380$ papers refer to the Ising model every year.
The {\em Google} search engine indicates over 279 thousands web pages which contain ``Ising model'' phrase.

The beauty and the popularity of this model lies in both its simplicity and possible applications from pure and applied physics, via life sciences to social sciences.
In the way similar to the percolation phenomenon, the Ising model is one of the most convenient way of numerical investigations of second order phase transitions.

In the simplest case, the Ising model may be used to simulate the system of interacting spins which are placed at the nodes of graphs or regular lattices.
In its basic version only two values of the spin variable are available, i.e. $S=-\frac{1}{2}$ and $S=+\frac{1}{2}$.
This is the classical Ising $S=\frac{1}{2}$ model. 
For a square lattice this model defines the universality class of phase transitions with analytically known critical exponents which describe the system behaviour near the critical point.
The critical point separates two --- ordered and disordered --- phases.

One of possible generalisation of the Ising model is to enlarge the set of possible spin values (like in the Potts model \cite{potts1,potts2}).
The Ising $S=1$ model corresponds to three possible spin values, i.e. $S\in \{-1,0,+1\}$, Ising $S=\frac{3}{2}$ allows for four spin variables $S\in\{\pm \frac{3}{2},\pm\frac{1}{2} \}$, {\em etc}.
The Ising $S\ne\frac{1}{2}$ model on various networks and lattices may form universality classes other than the classical square lattice Ising model.

The spin models for $S=1$ were extensively studied by several approximate techniques in two and three dimensions and their phase diagrams are well known \cite{BC1,BC2,BC3,BC4,BC5,BC6,BC7}.
The case $S>1$ has also been investigated according to several procedures \cite{BC8,BC9,BC10,BC11,BC12,BC13,landaupla}. 
The Ising model $S=1$ on directed Barab\'asi--Albert network was studied by Lima in 2006 \cite{lima-2006}. It was shown, that the system exhibits first-order phase transition. The result is qualitatively different from the results for this model on a square lattice, where a second-order phase transition is observed.

\begin{figure*}[!hbt]
\begin{center}
\includegraphics[clip, viewport=50 64 641 395, width=.35\textwidth]{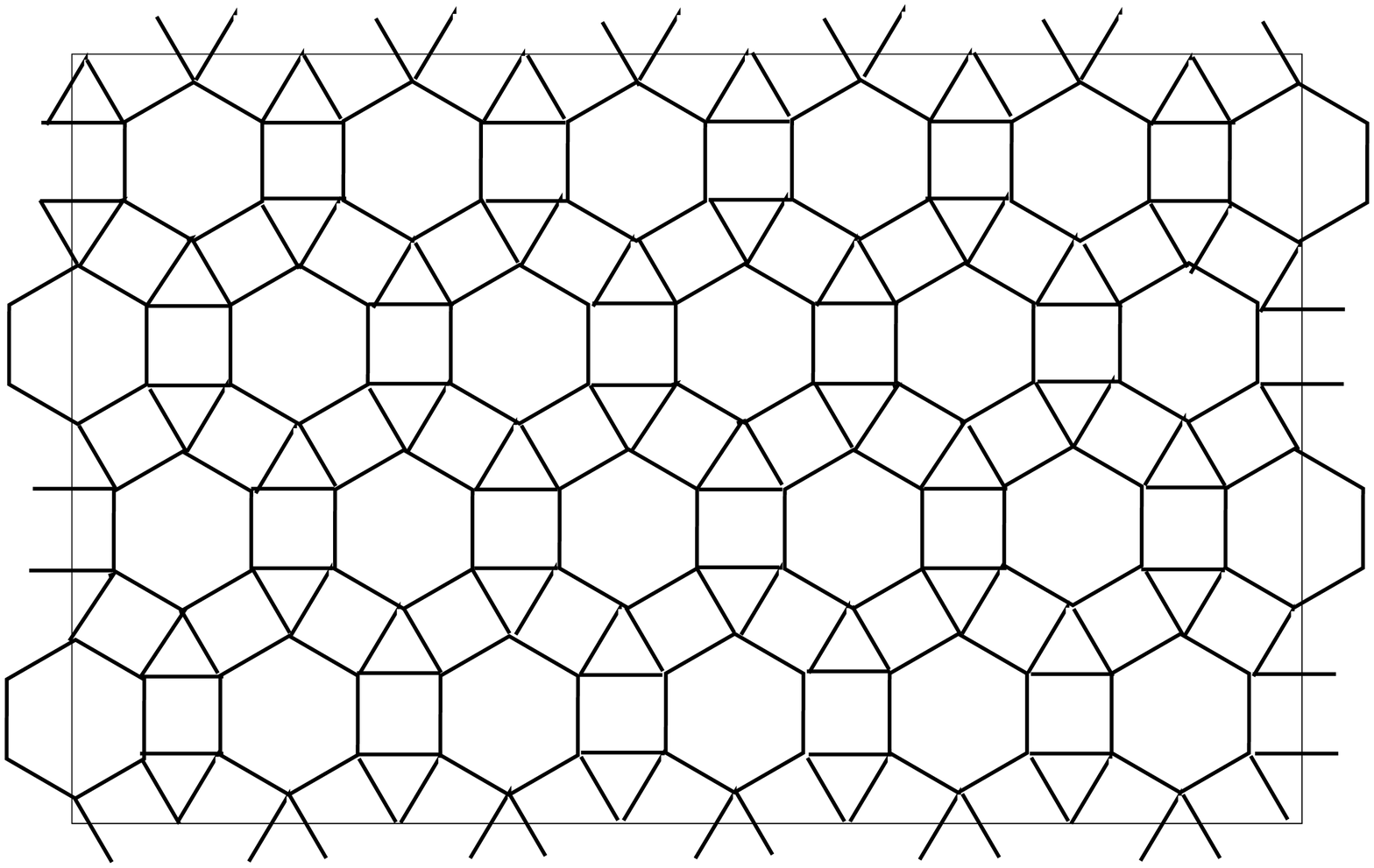}
\hspace{1cm} \includegraphics[clip, viewport=50 64 641 395, width=.35\textwidth]{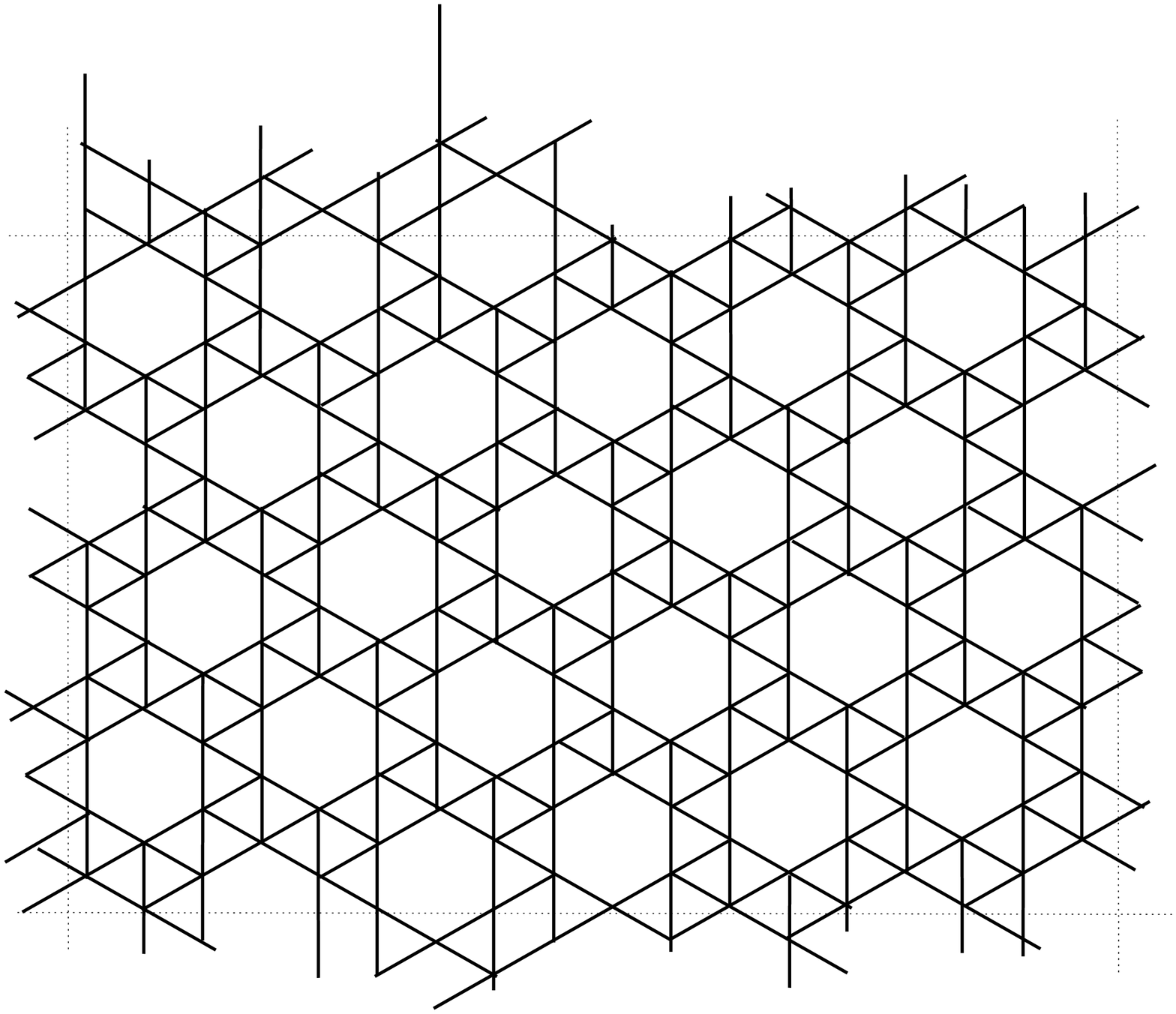}
\end{center}
\caption{\label{fig-AL} Topology of $(3,4,6,4)$ [left] and $(3^4,6)$ [right] AL.}
\end{figure*}

In this paper we study the Ising $S=1$ model on two Archimedean lattices (AL), namely on $(3,4,6,4)$ and $(3^4,6)$. The topologies of $(3,4,6,4)$ and $(3^4,6)$ AL are presented in Fig. \ref{fig-AL}. Critical properties of these lattices were investigated in terms of site percolation in Ref. \cite{suding-1999}. Topologies of all eleven existing AL are given there as well. Also the critical temperatures for Ising $S=\frac{1}{2}$ model \cite{malarz-2005} and voter model \cite{lima-malarz} on those AL were estimated numerically.

Here, with extensive Monte Carlo simulations we show that the Ising $S=1$ model on $(3,4,6,4)$ and $(3^4,6)$ AL exhibits a second-order phase transition with critical exponents that {\em do not} fall into universality class of the square lattice Ising $S=\frac{1}{2}$ model.

\section{Model and simulation}

We consider the two-dimensional Ising $S=1$ model on $(3,4,6,4)$ and $(3^4,6)$ AL lattices.
The Hamiltonian of the system can be written as
\begin{equation}
\label{hamiltonian}
{\cal H}=-J\sum_{i=1}^N \sum_{j>i}^N S_iS_j,
\end{equation}
where spin variable $S_i$ takes values $-1$, $0$, $+1$ and decorates every $N=6L^2$ vertex of the AL.
In Eq. \eqref{hamiltonian} $J$ is the magnetic exchange coupling parameter.

The simulations have been performed for different lattice sizes $L=8$, 16, 32, 64 and 128.
For each system with $N=6L^2$ spins and given temperature $T$ we performed Monte Carlo simulation in order to evaluate the system magnetisation $m$.
The simulations start with a uniform configuration of spins ($S_i=+1$, but the results are independent on the initial configuration). 
It takes $10^5$ Monte Carlo steps (MCS) per spin for reaching the steady state, and then the time average over the next $10^5$ MCS are estimated. One MCS is accomplished when all $N$ spins are investigated whether they should flip or not. We carried out $N_{\text{run}}=20$ to $50$ independent simulations for each lattice and for given set of parameters $(N,T)$.
We have employed the heat bath algorithm for the spin dynamic.

We evaluate the average magnetisation $M$, the susceptibility $\chi$, and the magnetic 4-th order
cumulant $U$:
\begin{subequations}
\label{eq-mchiU}
\begin{equation}
\label{eq-m}
 M(T,L)=\langle |m|\rangle,
\end{equation}
\begin{equation}
\label{eq-chi}
\dfrac{k_BT}{J}\cdot\chi(T,L)=N(\langle m^2\rangle- \langle|m|\rangle^2),
\end{equation}
\begin{equation}
\label{eq-U}
 U(T,L)=1-\frac{\langle m^4\rangle}{3 \langle|m|\rangle^2},
\end{equation}
\end{subequations}
where $m=\sum_iS_i/N$ and $k_B$ is the Boltzmann constant.
In the above equations $\langle ...\rangle$ stands for thermodynamic average.

In the infinite-volume limit these quantities \eqref{eq-mchiU} exhibit singularities at the transition point $T_C$.
In finite systems the singularities are smeared out and scale in the critical region according to
\begin{subequations}
\label{ffs}
\begin{equation}
\label{ffs-beta}
 M=L^{-\beta/\nu}f_{M}(x),
\end{equation}
\begin{equation}
\label{ffs-gamma}
 \chi=L^{-\gamma/\nu}f_{\chi}(x),
\end{equation}
\end{subequations}
where $\nu$, $\beta$ and $\gamma$ are the usual critical exponents, and $f_{i}(x)$ are finite size scaling (FSS) functions with $ x=(T-T_{C})L^{1/\nu}$
being the scaling variable. 
Therefore, from the size dependence of $M$ and $\chi$ one can obtain the exponents $\beta/\nu$ and $\gamma/\nu$, respectively.

The maximum value of susceptibility also scales as $L^{\gamma/\nu}$.
Moreover, the value of temperature $T^*$ for which $\chi$ has a maximum, 
is expected to scale with the system size as
\begin{equation}
\label{ffs-nu}
T^*(L)=T_C+bL^{-1/\nu},
\end{equation}
where the constant $b$ is close to unity \cite{binder-2005}. Therefore, the Eq. $\eqref{ffs-nu}$
may be used to determine the exponent $1/\nu$. We have checked also if the calculated
exponents satisfy the hyper-scaling hypothesis
\begin{equation}
\label{eq-deff}
2\beta/\nu+\gamma/\nu=D_{\text{eff}}
\end{equation}
in order to get the effective dimensionality, $D_{\text{eff}}$, for both investigated AL lattices.

\begin{figure*}[hbt]
\psfrag{m}{$M$}
\psfrag{K}{$k_BT/J$}
\psfrag{L=}{$L=$}
\psfrag{3464}{$(3,4,6,4)$}
\psfrag{33336}{$(3^4,6)$}
\begin{center}
\includegraphics[width=0.48\textwidth]{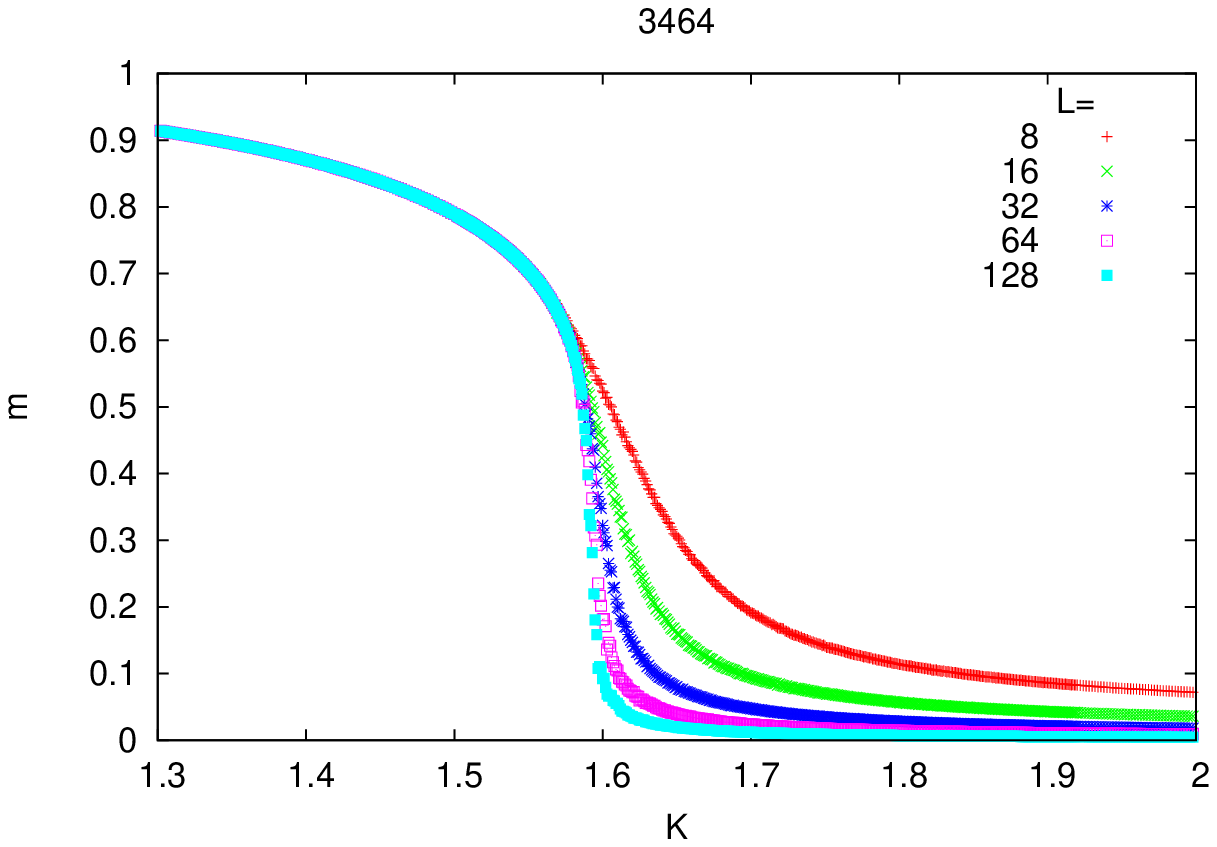}
\includegraphics[width=0.48\textwidth]{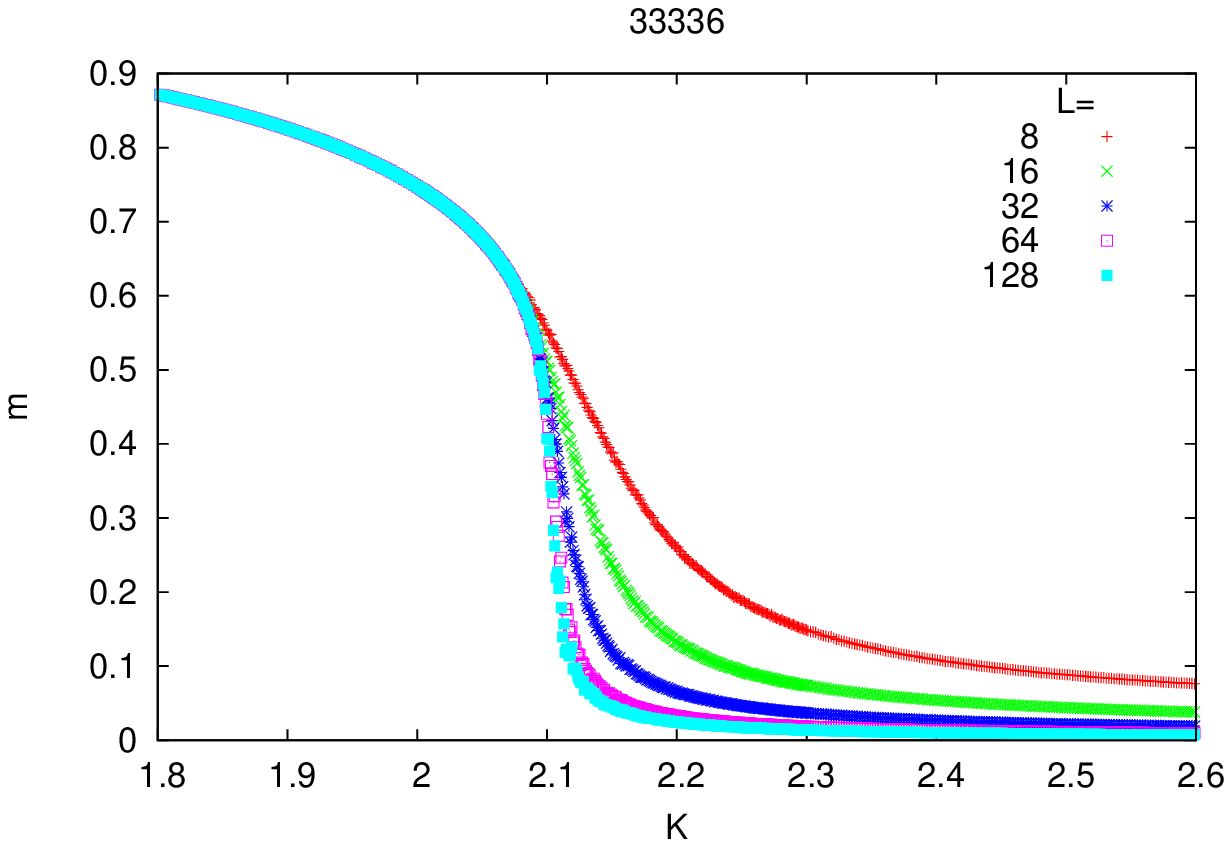}
\end{center}
\caption{The magnetisation $M$ as a function of the temperature $T$, for $L=8, 16, 32, 64$, and $128$ and for  $(3,4,6,4)$ and $(3^4,6)$ AL.}
\label{fig-m}
\end{figure*}

\begin{figure*}[hbt]
\psfrag{K}{$k_BT/J$}
\psfrag{chi}{$\chi$}
\psfrag{L=}{$L=$}
\psfrag{3464}{$(3,4,6,4)$}
\psfrag{33336}{$(3^4,6)$}
\begin{center}
\includegraphics[width=0.48\textwidth]{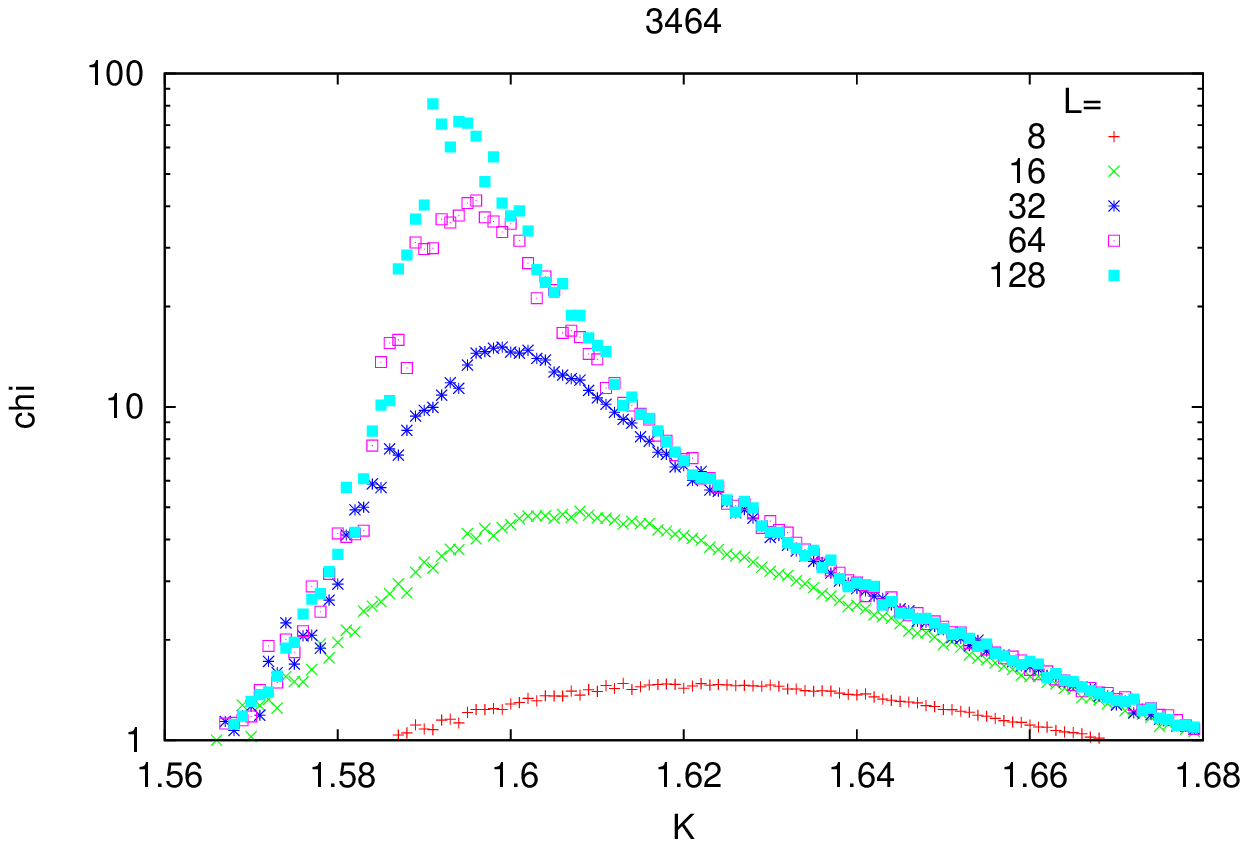}
\includegraphics[width=0.48\textwidth]{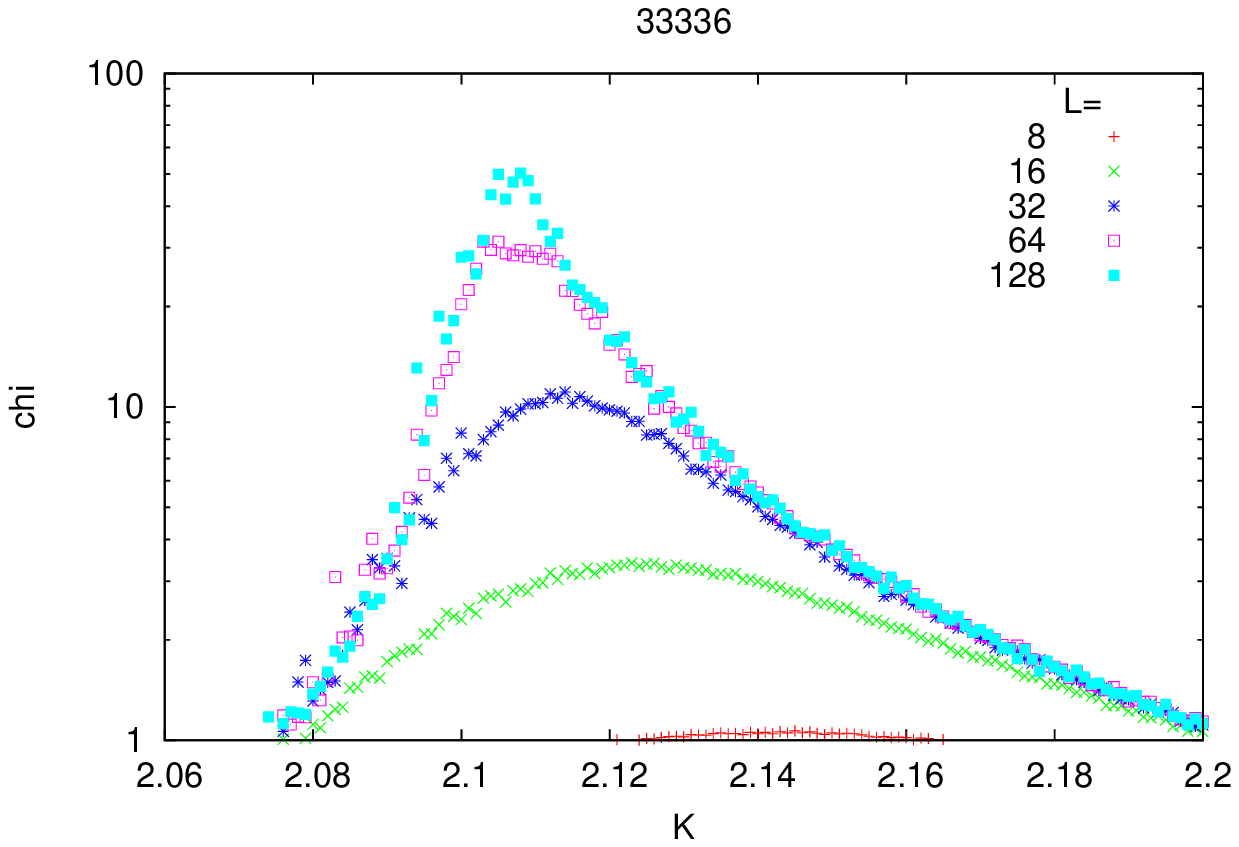}
\end{center}
\caption{The susceptibility $\chi$ versus temperature $T$, for $(3,4,6,4)$ and $(3^4,6)$ AL.}
\label{fig-chi}
\end{figure*}

\section{Results and discussion}

The dependence of the magnetisation $M$ on the temperature $T$, obtained from simulations on $(3,4,6,4)$ and $(3^4,6)$ AL with $N=6L^2$ ranging from $384$ to $98304$ sites is presented in Fig. \ref{fig-m}.
The shape of magnetisation curve versus temperature, for a given value of $N$, suggests the presents of the second-order transition phase in the system. The phase transition occurs at the critical value $T_C$ of temperature.

In order to estimate the critical temperature $T_C$ we calculate the fourth-order 
Binder cumulants given by Eq. \eqref{eq-U}. It is well known that these
quantities are independent of the system size at $T_C$ and should intercept there \cite{binder}.

In Fig. \ref{fig-chi} the corresponding behaviour of the susceptibility $\chi$ is presented.

\begin{figure*}[hbt]
\psfrag{K}{$k_BT/J$}
\psfrag{L=}{$L=$}
\psfrag{redU}{$-\ln(1-3U/2)$}
\psfrag{3464}{$(3,4,6,4)$}
\psfrag{33336}{$(3^4,6)$}
\begin{center}
\includegraphics[width=0.48\textwidth]{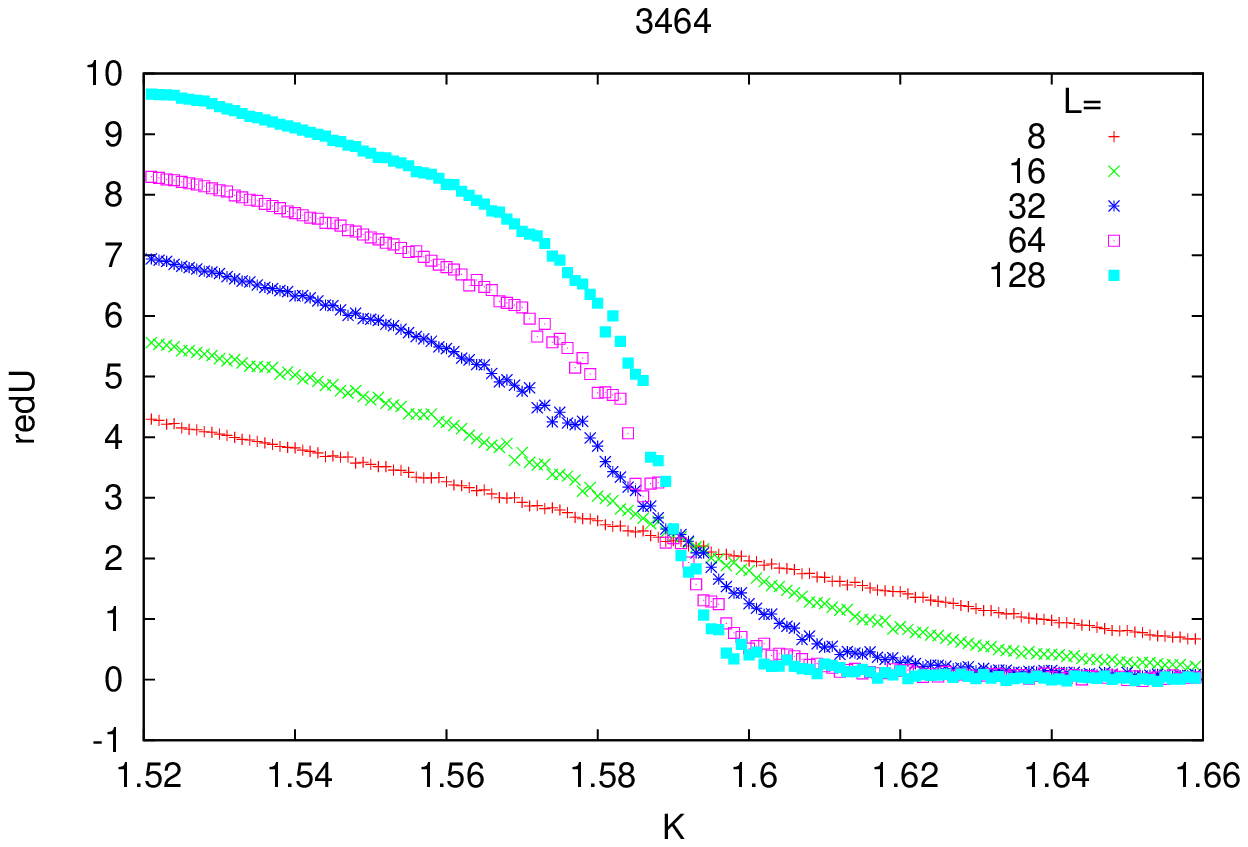}
\includegraphics[width=0.48\textwidth]{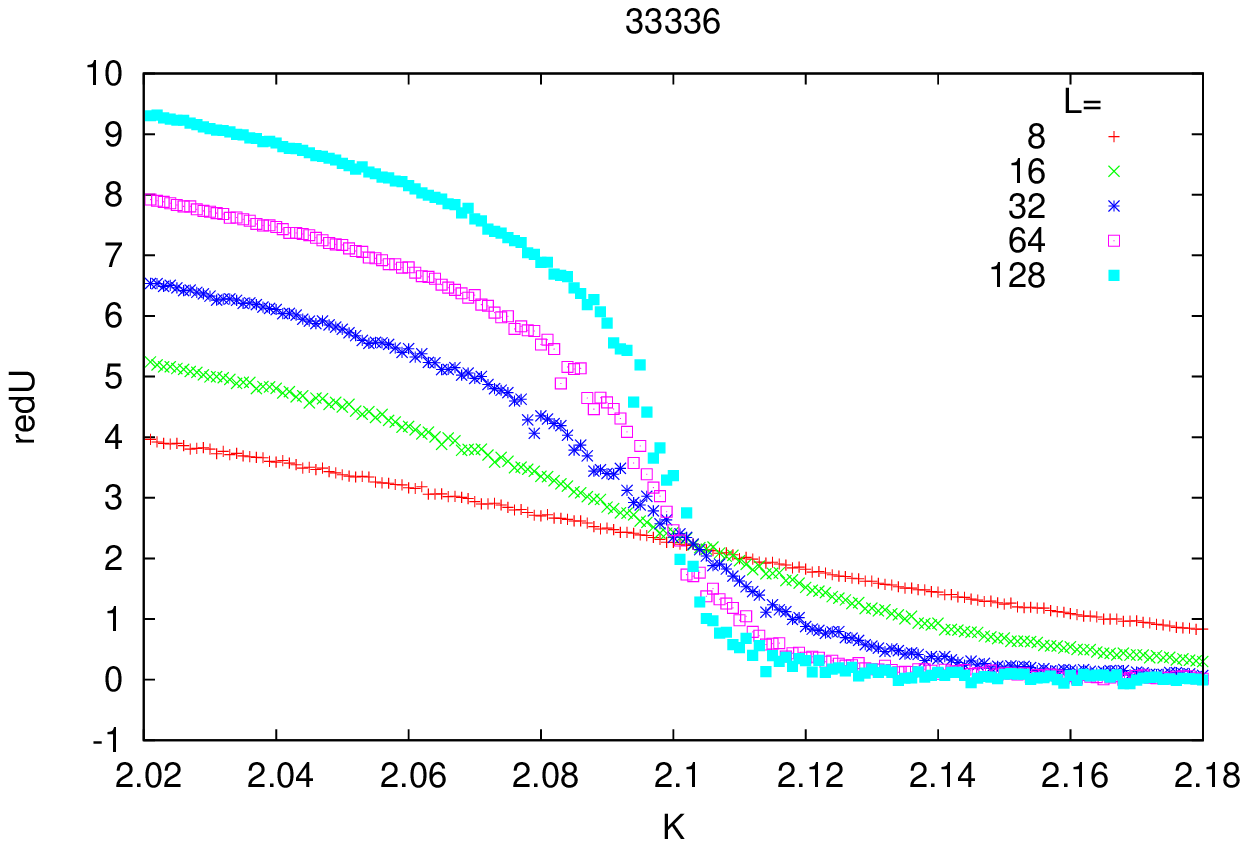}
\end{center}
\caption{The reduced Binder's fourth-order cumulant $U$ as a function of the temperature $T$, for $(3,4,6,4)$ and $(3^4,6)$ AL.}
\label{fig-U}
\end{figure*}

In Fig. 4 the fourth-order Binder cumulant is shown as a function of the temperature for several values of $L$. Taking two 
largest lattices (for $L=64$ and $L=128$) we have $T_C=1.590(3)$ and $T_C= 2.100(3)$ for $(3,4,6,4)$ and $(3^4,6)$ AL, respectively.

In order to go further in our analysis we also computed the modulus of the magnetisation
at the inflection $M^*=M(T_C)$.
The estimated exponents $\beta/\nu$ values are $0.180(20)$ and $0.103(7)$ for $(3,4,6,4)$ and $(3^4,6)$ AL, respectively.

Basing on the dependence $\ln\chi$ on $\ln L$ we estimated
$\gamma/\nu=1.46(8)$ and  $\gamma/\nu=1.44(8)$ for $(3,4,6,4)$ and $(3^4,6)$ AL, respectively.

To obtain the critical exponent $1/\nu$, we used the scaling relation \eqref{ffs-nu}. The calculated values of the exponents $1/\nu$ are $0.83(5)$ for $(3,4,6,4)$ and $1/\nu=0.94(5)$ for $(3^4,6)$. Eq. \eqref{eq-deff} yields effective dimensionality of the systems $D_{\text{eff}}=1.82(4)$ for $(3,4,6,4)$ and $D_{\text{eff}}=1.64(5)$ for $(3^4,6)$.

The above results, indicate that the Ising $S=1$ model on $(3,4,6,4)$ and $(3^4,6)$ AL {\em does not fall} in the same universality class as the square lattice Ising model, for which the critical exponents are known analytically i.e. $\beta=\frac{1}{8}=0.125$, $\gamma=\frac{7}{4}=1.75$ and $\nu=1$.
We have checked numerically, that Ising $S=\frac{1}{2}$ model reproduces these critical exponents with reasonable accuracy for both studied lattices \cite{mostowicz}.
We improved the value of the critical temperature $T_C$ for these two lattices and $S=\frac{1}{2}$ as well, with respect to Ref. \cite{malarz-2005}.

The results are collected in Tab. \ref{tab}.

\begin{table*}
\caption{\label{tab} Critical points and critical points exponents for $(3,4,6,4)$ and $(3^4,6)$ AL.
For comparison, the exact values for the square lattice Ising $S=\frac{1}{2}$ model are included as well.}
\begin{ruledtabular}
\begin{tabular}{ll lllll}
 & $S$ & $k_BT_C/J$ & $\beta/\nu$ & $\gamma/\nu$ & $1/\nu$ & $D_{\text{eff}}$ \\ \hline
$(3,4,6,4)$    & 1             & 1.590(3) & 0.180(20) & 1.46(8)  & 0.83(5)  & 1.82(4)  \\
$(3^4,6)$      & 1             & 2.100(3) & 0.103(8)  & 1.44(8)  & 0.94(5)  & 1.64(5)  \\ \hline
$(3,4,6,4)$    & $\frac{1}{2}$ & 2.145(3) & 0.123(17) & 1.680(74)& 1.066(44)& 1.926(84)\\
$(3^4,6)$      & $\frac{1}{2}$ & 2.784(3) & 0.113(10) & 1.726(8) & 1.25(13) & 1.952(22)\\ \hline
square $(4^4)$ & $\frac{1}{2}$ & $2/\text{arcsinh}(1)$ & $\frac{1}{8}$ & $\frac{7}{4}$ & 1 & 2 \\
\end{tabular}
\end{ruledtabular}
\end{table*}

Except the exponent $\nu$, all critical exponents for $S=1$ differ for more than three numerically estimated uncertainties from those given analytically.

\begin{acknowledgments}
FWSL acknowledges the Brazilian agency FAPEPI (Teresina, Piau\'{\i}, Brazil) for its financial support.
This work also was supported (in part) from the AGH-UST project 11.11.220.01.
Part of the calculations was carried out on {\em SGI Altix 1350} system at the computational park CENAPAD.UNICAMP-USP, S\~ao Paulo, Brazil and on {\em SGI Altix 3700} system at Academic Computer Centre Cyfronet-AGH in Cracow, Poland.
\end{acknowledgments}

\end{document}